\documentclass[aps,preprint,nofootinbib]{revtex4-1}
\usepackage{amsmath}
\usepackage{physics}
\usepackage{graphicx}
\usepackage{amssymb}
\usepackage[colorlinks,citecolor=red]{hyperref}

\usepackage{color}    
\usepackage{ulem}
\usepackage{multirow}

\usepackage{ulem}
\usepackage{makecell}

\newcommand{\bqa}{\begin{eqnarray}}
\newcommand{\eqa}{\end{eqnarray}}
\newcommand{\beq}{\begin{equation}}
\newcommand{\eeq}{\end{equation}}

\allowdisplaybreaks

\begin{document}

\title{Two-loop electroweak corrections to the Higgs boson rare decay process $H\to Z\gamma$}

\author{Zi-Qiang Chen}
\email{chenziqiang13@gzhu.edu.cn}
\affiliation{School of Physics and Materials Science, Guangzhou University, Guangzhou 510006, China}

\author{Long-Bin Chen}
\email{Corresponding author: chenlb@gzhu.edu.cn}
\affiliation{School of Physics and Materials Science, Guangzhou University, Guangzhou 510006, China}

\author{Cong-Feng Qiao}
\email{Corresponding author: qiaocf@ucas.ac.cn}
\affiliation{School of Physical Sciences,  University of Chinese Academy of Science,Yuquan Road 19A, Beijing 100049, China}

\author{Ruilin Zhu}
\email{Corresponding author: rlzhu@njnu.edu.cn}
\affiliation{Department of Physics and Institute of Theoretical Physics, Nanjing Normal University, Nanjing, Jiangsu 210023, China}

{\it Dedicated to the memory of P.W. Higgs (1929-2024)}

\begin{abstract}
Recently, the ATLAS and CMS collaborations jointly announced the first evidence of the rare Higgs boson decay channel $H\to Z\gamma$,
with a ratio of $2.2\pm 0.7$ times the leading order standard model (SM) prediction. 
In order to face this challenge, it is urgent to produce an even more accurate calculation within the SM. To this end, we calculate in this paper the next-to-leading order (NLO) electroweak (EW) corrections to the $H\to Z\gamma$ process, in which the NLO quantum chromodynamics (QCD) corrections were found tiny. 
Our calculation finds that the inclusion of NLO EW corrections greatly enhances the prediction reliability. 
To tame the theoretical uncertainty, we adopt five different renormalization schemes. Combining our result with previous NLO QCD corrections and the signal-background interference, we conclude that the excess in $H\to Z\gamma$ cannot be explained within the SM. 
In fact, the incompatibility between the SM prediction and the LHC measurement of the concerned process is exacerbated upon considering the higher order EW corrections, which implies that something beyond the SM could be involved. 
\end{abstract}
\maketitle

\section{Introduction}
The Higgs mechanism \cite{Higgs:1964pj,Englert:1964et} holds a paramount position within the standard model (SM) of particle physics, playing a decisive role in illuminating the fundamental underpinnings of mass generation in elementary particles. 
Since the discovery of the Higgs boson in 2012 \cite{ATLAS:2012yve,CMS:2012qbp}, the study of its properties is the top priority of the Large Hadron Collider (LHC) experiment.
Although most measurements up to now show no significant deviation from the SM  predictions \cite{ATLAS:2022vkf,CMS:2022dwd}, it is conceivable that much more detailed and precise investigations are required to verify the correctness of the SM and search for a new physics signal.
Recently, the ATLAS and CMS collaborations have teamed up to announce the first evidence for the Higgs boson decay to a $Z$ boson and a photon, with a statistical significance of $3.4\sigma$ \cite{ATLAS:2023yqk}. 
The measured signal yield is $\mu = 2.2\pm 0.7$ times the SM prediction.
 The interpretation of this signal is of great interest due to its potential excess.

As a loop-induced process, the $H\to Z\gamma$ decay is inherently sensitive to potential deviation or extension of the SM \cite{Low:2011gn,Low:2012rj,Azatov:2013ura,Cao:2018cms,Boto:2023bpg}.
To discern the potential new physics signal with heightened clarity, it is important to acquire the most accurate available prediction of the SM.
The leading order (LO) calculation of this channel was achieved long ago \cite{Cahn:1978nz,Bergstrom:1985hp}, with an uncertainty of about $10\%$ for $125\; \text{GeV}$ SM Higgs boson \cite{LHCHiggsCrossSectionWorkingGroup:2013rie}; the uncertainty of this channel is relatively larger compared with other decay channels \cite{ParticleDataGroup:2022pth}.
The two-loop quantum chromodynamics (QCD) correction to $H \to Z \gamma$ decay was investigated in Refs. \cite{Spira:1991tj,Gehrmann:2015dua,Bonciani:2015eua}.
According to the results, the QCD corrections turned out to be rather small, which is about $0.22\%$ relative to the LO contribution.
The authors of Ref. \cite{Buccioni:2023qnt} investigated the interference effect between the $gg\to H\to Z\gamma$ signal and the $gg\to Z\gamma$ background at next-to-leading order (NLO) QCD accuracy.
They found that the interference has a destructive impact on the total rate of $\mathcal{O}(-3\%)$.
All these corrections are small in comparison with current experimental accuracy, and the tension between the SM prediction and experimental measurement are not reduced.

At LO, the relative contribution of the $W$-boson loop is about $110\%$ (the interference between the $W$-boson loop and top-quark loop is destructive), hence the two-loop electroweak (EW) corrections can be more significant than the QCD counterparts.
Although the two-loop EW corrections to $H\to \gamma\gamma$ and $H\to gg$ processes were carried out 16 years ago,  \cite{Actis:2008ts}, the EW corrections to $H\to Z\gamma$ have remained elusive until now.
To interpret the experimental signal properly and improve the accuracy of the SM prediction, in this work, we calculate the NLO EW corrections to the decay of a Higgs boson into a $Z$ boson and a photon.

\section{The NLO EW corrections to $H \to Z \gamma$ decay}

\begin{figure}[t]
\includegraphics[width=0.9\textwidth]{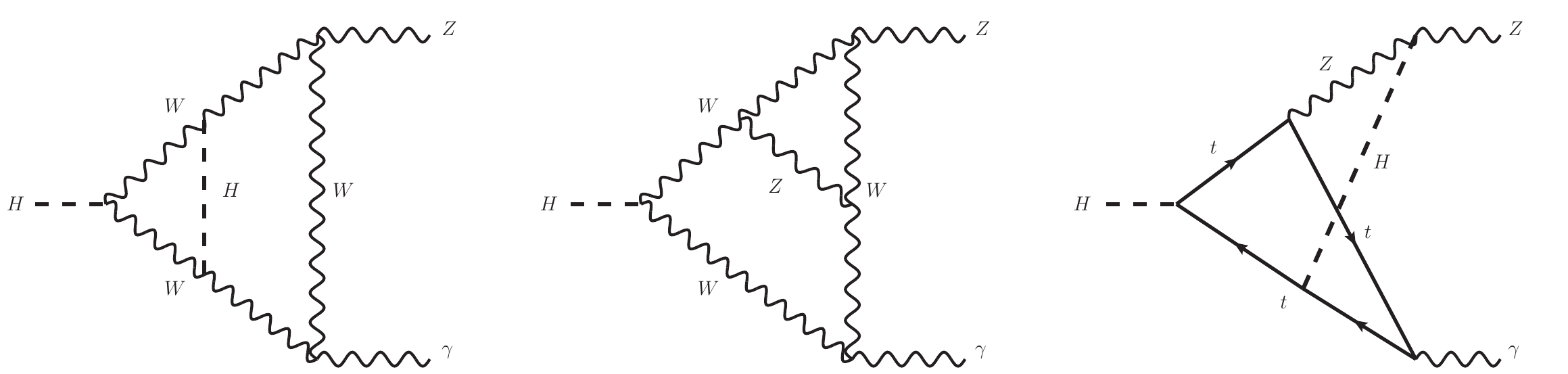}
\caption{Sample  Feynman diagrams for NLO EW corrections to $H\to Z\gamma$ decay.}
\label{fig:feynman-diagram}
\end{figure}

The calculation is carried out in the Feynman gauge.
The conventional dimensional regularization with $D=4-2\epsilon$ is adopted to regularize the ultraviolet divergences.
The masses of light fermions (leptons and quarks except the top) are neglected whenever possible; they are only taken into account in the mass-singular logarithms originating from the photonic vacuum polarization.

The NLO EW corrections comprise about 5000 two-loop Feynman diagrams; three samples of them are shown in Fig.~\ref{fig:feynman-diagram}.
We use the package FEYNARTS \cite{Hahn:2000kx} to generate the diagrams as well as the amplitudes.
With the aid of FEYNCALC \cite{Shtabovenko:2020gxv}, the amplitudes are then expressed in terms of scalar integrals.
The package FIRE \cite{Smirnov:2019qkx} is employed to perform the integration-by-parts (IBP) reduction \cite{Chetyrkin:1981qh},
and the master integrals are evaluated numerically using the AMFLOW \cite{Liu:2022chg} package.
As a cross-check, some master integrals are also evaluated using either analytical approach or the FIESTA \cite{Smirnov:2021rhf} package, and perfect agreements have been found. 

The amplitudes we obtained are free of infrared and collinear divergences; only $1/\epsilon$ pole appears.  The renormalization is implemented according to Ref. \cite{Denner:1991kt}.
After including the counterterm diagrams, all ultraviolet divergences, which regularized as  a $1/\epsilon$ pole, are eliminated as expected.
Unlike the ultraviolet divergences, the $\ln (m_f)$ terms, where $m_f$ denotes the mass of light fermion, may not cancel, depending on the scheme chosen for the electromagnetic coupling constant $\alpha$.
As an estimation of theoretical uncertainty, here we use five different schemes:
\begin{enumerate}
\item $\alpha(0)$ scheme: all couplings are set equal to the fine-structure constant $\alpha(0)$; 
\item $\alpha(m_Z^2)$ scheme: all couplings are set equal to the effective value $\alpha(m_Z^2)$; 
\item $G_\mu$ scheme: all couplings are derived from the Fermi constant by using $\alpha_{G_\mu}=\sqrt{2}G_\mu m_W^2(1-m_W^2/m_Z^2)$;
\item mixed scheme 1: the coupling relative to the real external photon is taken as $\alpha(0)$, while other couplings are taken as $\alpha(m_Z^2)$;
\item mixed scheme 2: the coupling relative to the real external photon is taken as $\alpha(0)$, while other couplings are taken as $\alpha_{G_\mu}$.
\end{enumerate}
The cancellation of $\ln (m_f)$ terms only occurs in the last two mixed schemes.
Note, to avoid double counting and to keep gauge invariance, the charge renormalization constants corresponding to $\alpha(m_Z^2)$ and $\alpha_{G_\mu}$ are redefined as $\delta Z_e^{\alpha(m_Z^2)} =\delta Z_e-\Delta \alpha(m_Z^2)/2$, $\delta Z_e^{G_\mu} =\delta Z_e-\Delta r^{(1)}/2$, where $\delta Z_e$ is the renormalization constant corresponding to $\alpha(0)$.
The details about the quantities $\Delta \alpha(m_Z^2)$ and $\Delta r^{(1)}$ can be found in Ref. \cite{Denner:2019vbn}.

\section{Numerical results}
We employ the following ensemble of input parameters
\begin{align}
& \alpha(0)=1/137.036,\quad\quad \alpha(m_Z^2)=1/127.951, \quad\quad G_\mu=1.16638\times 10^{-5},\nonumber \\
&m_H=125.25\; {\rm GeV}, \quad\quad m_Z=91.19\; {\rm GeV},\quad\quad m_W=80.38\; {\rm GeV},\nonumber\\
&m_u=66\; {\rm MeV},\quad\quad m_c=1.2\; {\rm GeV},\quad\quad  m_t=172.69\; {\rm GeV}, \nonumber \\
&m_d=66\; {\rm MeV},\quad\quad m_s=150\; {\rm MeV},\quad\quad  m_b=4.6\; {\rm GeV}, \nonumber \\
&m_e=0.511\; {\rm MeV},\quad\quad m_\mu=105.66\; {\rm MeV},\quad\quad m_\tau=1.77686\; {\rm GeV}.
\end{align}
Here, the masses of the light quarks are adjusted to reproduce the hadronic contribution to the photonic vacuum polarization, which takes the value $\Delta \alpha_{\rm had}^{(5)}(m_Z^2)=0.027730$ \cite{Jegerlehner:2001ca,Dittmaier:2009cr}; other parameters are taken from the Particle Data Group's Review of Particle Physics \cite{ParticleDataGroup:2022pth}.
The weak mixing angle is fixed by $c_W=m_W/m_Z$, $s_W=\sqrt{1-c_W^2}$.

Our main results are presented in Table \ref{numeric-values}.
There we list the LO decay widths $\Gamma^{\rm LO}$, the NLO decay widths $\Gamma^{\rm NLO}_{\rm EW}$, and the relative EW corrections $\delta_{\rm EW}$ of different schemes.
For clarity purposes, the input parameters of each scheme are listed as well. One can see that the NLO EW corrections are negative in all schemes except the $\alpha(0)$ scheme.
Notice that the mixed scheme 2 exhibits excellent convergence behavior with $\delta_{\rm EW}=-0.75\%$, since this scheme absorbs not only $\ln(m_f)$ terms but also some corrections to the $\rho$ parameter into the LO contributions \cite{Denner:2019vbn}.
Taking the result of mixed scheme 2 as the central value and the results of other schemes minus the central value as the theoretical uncertainty, we have $\Gamma^{\rm LO}=6.364^{+0.909}_{-0.444}\; \text{keV}$, $\Gamma^{\rm NLO}_{\rm EW}=6.316^{+0.027}_{-0.082}\; \text{keV}$.
It can be seen that the EW corrections greatly suppress the theoretical uncertainty, and enhance the prediction reliability.

\begin{table}[h]
\begin{tabular}{p{2cm}<{\centering} p{3cm}<{\centering} p{2cm}<{\centering} p{2cm}<{\centering} p{2cm}<{\centering}}
\hline
scheme & input parameters & $\Gamma^{\rm LO}$ (keV) & $\Gamma^{\rm NLO}_{\rm EW}$ (keV) & $\delta_{\rm EW}$ ($\%$)  \\
\hline
$\alpha(0)$ & $\alpha(0)$, $m_f$ & 5.920 & 6.234 & 5.3  \\
$\alpha(m_Z^2)$ & $\alpha(m_Z^2)$, $m_f$ & 7.273 & 6.303 & $-13$ \\
$G_\mu$ & $G_\mu$, $m_f$ & 6.599 & 6.343 & $-3.9$ \\
mixed 1 & $\alpha(0)$, $\alpha(m_Z^2)$ & 6.791 & 6.316 & $-7.0$ \\
mixed 2 & $\alpha(0)$, $G_\mu$ & 6.364 & 6.316 & $-0.75$ \\
\hline
\end{tabular}
\caption{The LO and NLO decay widths of $H\to Z\gamma$ under different coupling schemes. The relative EW corrections are also given.}
\label{numeric-values}
\end{table}

\begin{figure}[t]
    \centering
    \includegraphics[width=0.8\textwidth]{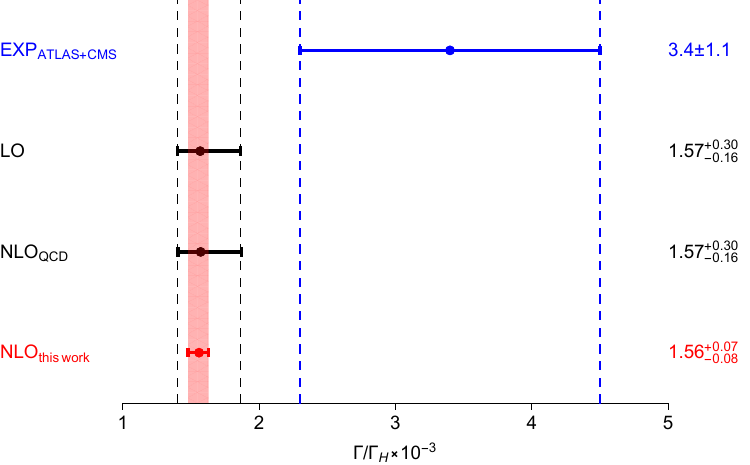}
    \caption{Theoretical predictions in comparison with experimental data.  ``$\mathrm{EXP_{ATLAS+CMS}}$'' denotes the latest data from the ATLAS and CMS collaborations ~\cite{ATLAS:2023yqk}. ``LO'', ``$\mathrm{NLO_{QCD}}$'', ``$\mathrm{NLO_{this\; work}}$'' denote the theoretical predictions at leading order, next-to-leading order including QCD corrections only, next-to-leading order including both QCD  and EW corrections, respectively. Note, all these theoretical results include the $b$-mass corrections. The total decay width of the Higgs boson is taken as $\Gamma_H=4.07\pm 0.16\; \text{MeV}$ \cite{ParticleDataGroup:2022pth}.}
    \label{fig:spectrum}
\end{figure}

To make a more precise prediction, it is necessary to include other effects that are comparable to NLO EW corrections.
We investigate the bottom quark mass effect and find $\delta_b=(\Gamma^{\rm LO}_{\text{with}\ b}-\Gamma^{\rm LO})/\Gamma^{\rm LO}=0.28\%$.
The NLO QCD corrections to $H\to Z\gamma$ were calculated in Refs. \cite{Gehrmann:2015dua,Bonciani:2015eua}, and according to their results, the relative QCD correction $\delta_{\rm QCD}$ with respect to $\Gamma^{\rm LO}_{\text{with}\ b}$ is about $0.22\%$.
Combining all these contributions, we have
\begin{align}
\Gamma^{\rm NLO}&=\Gamma^{\rm NLO}_{\rm EW}+\Gamma^{\rm LO}(1+\delta_b)(1+\delta_{\rm QCD})-\Gamma^{\rm LO}\nonumber\\
&=6.348^{+0.028}_{-0.085}\; \text{keV}.
\end{align}
As the SM prediction for the total decay width of the Higgs boson is $\Gamma_H=4.07\pm 0.16\; \text{MeV}$ \cite{ParticleDataGroup:2022pth}, we finally obtain $\text{Br}(H\to Z\gamma)=1.56^{+0.07}_{-0.08}\times 10^{-3}$.
This result does not modify the signal strength $\mu=2.2\pm0.7$ measured by the ATLAS and CMS collaborations. In Fig.~\ref{fig:spectrum}, we provide a direct visual juxtaposition of the theoretical predictions for the $H\to Z\gamma$ decay alongside the corresponding experimental measurements. Notably, upon examining the figure, it becomes apparent that the incorporation of two-loop electroweak corrections has served to exacerbate the disparity between the theoretical predictions and the experimental data.

Finally, regarding the treatment of unstable particles, we would like to highlight two points:
\begin{itemize}
\item Strictly speaking, in standard perturbation theory which requires stable asymptotic in/out states, an unstable particle should only appear as an intermediate state.
Hence a more coherent treatment is to consider the full $H\to \ell \bar{\ell}\gamma$ process, \footnote{The Higgs boson is expected to be very narrow in comparison with the $Z$ boson.} including both the $Z$-resonant and nonresonant diagrams, with the $W$ and $Z$ masses renormalized in the complex-mass (CM) scheme \cite{Denner:2019vbn,Denner:2006ic,Actis:2006rc,Frederix:2018nkq}.
Further investigation on this topic is indispensable for a comprehensive signal-background analysis,
while the complete NLO calculation of the full process is very challenging. 
\item Although the decay width $\Gamma(H\to Z\gamma)$ does not have a precise meaning in standard perturbation theory, it can be extracted from the $H\to \ell \bar{\ell}\gamma$ signal by using the pole scheme \cite{Denner:2019vbn}.
In this work, in the calculation of $\Gamma(H\to Z\gamma)$, the $W$ and $Z$ bosons are treated as stable particles, with their masses renormalized in the on-shell (OS) scheme.
This treatment guarantees gauge invariance at NLO, 
while at next-to-next-to-leading order the OS scheme involves gauge dependent terms \cite{Sirlin:1991fd}.
A potential solution is to use the CM scheme with complex kinematical variables.
This approach, however, violates unitarity and introduces subtleties in analytical continuation \cite{Passarino:2010qk}.
On the other hand, for the $H\to Z\gamma$ process, the transition from the OS scheme to the CM scheme only causes corrections of order $\mathcal{O}(\Gamma_Z^2/m_Z^2)\sim \mathcal{O}(\alpha^2)$. \footnote{Considering the LO amplitude, which is purely real under the OS scheme. The CM scheme requires to identify the gauge boson masses with the complex pole values, i.e. $m_Z^2\to \mu_Z^2=m_Z^2+\mathcal{O}(\Gamma^2_Z/m_Z^2)+i\mathcal{O}(\Gamma_Z/m_Z)$,  $m_W^2\to \mu_W^2=m_W^2+\mathcal{O}(\Gamma^2_W/m_W^2)+i\mathcal{O}(\Gamma_W/m_W)$ \cite{Denner:2019vbn}. This leads to $\mathcal{M}_\text{LO}^\text{OS}\to \mathcal{M}_\text{LO}^\text{CM}=\mathcal{M}_\text{LO}^\text{OS}+\mathcal{O}(\Gamma^2_Z/m_Z^2)+i\mathcal{O}(\Gamma_Z/m_Z)$, and eventually $\left|\mathcal{M}_\text{LO}^\text{OS}\right|^2\to \left|\mathcal{M}_\text{LO}^\text{CM}\right|^2=\left|\mathcal{M}_\text{LO}^\text{OS}\right|^2+\mathcal{O}(\Gamma^2_Z/m_Z^2)$.}
\end{itemize}

\section{Conclusion}
In this paper, we investigated the Higgs decays into a photon and a $Z$ boson process at the NLO EW accuracy.
Numerical results indicate that the NLO EW corrections greatly suppress the theoretical uncertainty and enhance the prediction reliability.
By including the bottom quark mass effect and the NLO QCD corrections, we obtain the currently most precise SM prediction of the branching ratio $\text{Br}(H\to Z\gamma)=1.56^{+0.07}_{-0.08}\times 10^{-3}$.
This result, however, cannot reduce the tension between the SM prediction and experimental measurement.
We conclude that the excess in $H\to Z \gamma$ decays at ATLAS and CMS cannot be explained by including the higher order corrections within the SM. 
This could probably be attributed to the underestimated experimental uncertainties or the new physics beyond the SM.

\vspace{5cm} {\bf Acknowledgments}

We would like to thank Zhao-Huan Yu and Chen Zhou for useful discussions. 
This work was supported in part by the National Natural Science Foundation of China(NSFC) under the Grants No. 12075124, No. 12175048, No. 12235008, No. 12205061, and No. 12322503.
The work of L.B.C. was also supported by the Natural Science Foundation of Guangdong Province under Grant No. 2022A1515010041. 

\vspace{1.0cm}


\begin{thebibliography}{9}
\bibitem{Higgs:1964pj}
P.~W.~Higgs,
Phys. Rev. Lett. \textbf{13}, 508-509 (1964)

\bibitem{Englert:1964et}
F.~Englert and R.~Brout,
Phys. Rev. Lett. \textbf{13}, 321-323 (1964)

\bibitem{ATLAS:2012yve}
G.~Aad \textit{et al.} [ATLAS],
Phys. Lett. B \textbf{716}, 1-29 (2012)
[arXiv:1207.7214 [hep-ex]].

\bibitem{CMS:2012qbp}
S.~Chatrchyan \textit{et al.} [CMS],
Phys. Lett. B \textbf{716}, 30-61 (2012)
[arXiv:1207.7235 [hep-ex]].

\bibitem{ATLAS:2022vkf}
G.~Aad \textit{et al.} [ATLAS],
Nature \textbf{607}, no.7917, 52-59 (2022)
[erratum: Nature \textbf{612}, no.7941, E24 (2022)]
[arXiv:2207.00092 [hep-ex]].

\bibitem{CMS:2022dwd}
A.~Tumasyan \textit{et al.} [CMS],
Nature \textbf{607}, no.7917, 60-68 (2022)
[erratum: Nature \textbf{623}, no.7985, E4 (2023)]
[arXiv:2207.00043 [hep-ex]].

\bibitem{ATLAS:2023yqk}
G.~Aad \textit{et al.} [ATLAS and CMS],
Phys. Rev. Lett. \textbf{132}, no.2, 021803 (2024)
[arXiv:2309.03501 [hep-ex]].

\bibitem{Low:2011gn}
I.~Low, J.~Lykken and G.~Shaughnessy,
Phys. Rev. D \textbf{84}, 035027 (2011)
[arXiv:1105.4587 [hep-ph]].

\bibitem{Low:2012rj}
I.~Low, J.~Lykken and G.~Shaughnessy,
Phys. Rev. D \textbf{86}, 093012 (2012)
[arXiv:1207.1093 [hep-ph]].

\bibitem{Azatov:2013ura}
A.~Azatov, R.~Contino, A.~Di Iura and J.~Galloway,
Phys. Rev. D \textbf{88}, no.7, 075019 (2013)
[arXiv:1308.2676 [hep-ph]].

\bibitem{Cao:2018cms}
Q.~H.~Cao, L.~X.~Xu, B.~Yan and S.~H.~Zhu,
Phys. Lett. B \textbf{789}, 233-237 (2019)
[arXiv:1810.07661 [hep-ph]].

\bibitem{Boto:2023bpg}
R.~Boto, D.~Das, J.~C.~Romao, I.~Saha and J.~P.~Silva,
Phys. Rev. D \textbf{109}, no.9, 095002 (2024)
[arXiv:2312.13050 [hep-ph]].

\bibitem{Cahn:1978nz}
R.~N.~Cahn, M.~S.~Chanowitz and N.~Fleishon,
Phys. Lett. B \textbf{82}, 113-116 (1979)

\bibitem{Bergstrom:1985hp}
L.~Bergstrom and G.~Hulth,
Nucl. Phys. B \textbf{259}, 137-155 (1985)
[erratum: Nucl. Phys. B \textbf{276}, 744-744 (1986)]

\bibitem{LHCHiggsCrossSectionWorkingGroup:2013rie}
S.~Heinemeyer \textit{et al.} [LHC Higgs Cross Section Working Group],
[arXiv:1307.1347 [hep-ph]].

\bibitem{ParticleDataGroup:2022pth}
R.~L.~Workman \textit{et al.} [Particle Data Group],
PTEP \textbf{2022}, 083C01 (2022)

\bibitem{Spira:1991tj}
M.~Spira, A.~Djouadi and P.~M.~Zerwas,
Phys. Lett. B \textbf{276}, 350-353 (1992)

\bibitem{Gehrmann:2015dua}
T.~Gehrmann, S.~Guns and D.~Kara,
JHEP \textbf{09}, 038 (2015)
[arXiv:1505.00561 [hep-ph]].

\bibitem{Bonciani:2015eua}
R.~Bonciani, V.~Del Duca, H.~Frellesvig, J.~M.~Henn, F.~Moriello and V.~A.~Smirnov,
JHEP \textbf{08}, 108 (2015)
[arXiv:1505.00567 [hep-ph]].

\bibitem{Buccioni:2023qnt}
F.~Buccioni, F.~Devoto, A.~Djouadi, J.~Ellis, J.~Quevillon and L.~Tancredi,
Phys. Lett. B \textbf{851}, 138596 (2024)
[arXiv:2312.12384 [hep-ph]].

\bibitem{Actis:2008ts}
S.~Actis, G.~Passarino, C.~Sturm and S.~Uccirati,
Nucl. Phys. B \textbf{811}, 182-273 (2009)
[arXiv:0809.3667 [hep-ph]].

\bibitem{Hahn:2000kx}
T.~Hahn,
Comput. Phys. Commun. \textbf{140}, 418-431 (2001)
[arXiv:hep-ph/0012260 [hep-ph]].

\bibitem{Shtabovenko:2020gxv}
V.~Shtabovenko, R.~Mertig and F.~Orellana,
Comput. Phys. Commun. \textbf{256}, 107478 (2020)
[arXiv:2001.04407 [hep-ph]].

\bibitem{Smirnov:2019qkx}
A.~V.~Smirnov and F.~S.~Chuharev,
Comput. Phys. Commun. \textbf{247 }, 106877 (2020)
[arXiv:1901.07808 [hep-ph]].

\bibitem{Chetyrkin:1981qh}
K.~G.~Chetyrkin and F.~V.~Tkachov,
Nucl. Phys. B \textbf{192}, 159-204 (1981)

\bibitem{Liu:2022chg}
X.~Liu and Y.~Q.~Ma,
Comput. Phys. Commun. \textbf{283}, 108565 (2023)
[arXiv:2201.11669 [hep-ph]].

\bibitem{Smirnov:2021rhf}
A.~V.~Smirnov, N.~D.~Shapurov and L.~I.~Vysotsky,
Comput. Phys. Commun. \textbf{277}, 108386 (2022)
[arXiv:2110.11660 [hep-ph]].

\bibitem{Denner:1991kt}
A.~Denner,
Fortsch. Phys. \textbf{41}, 307-420 (1993)
[arXiv:0709.1075 [hep-ph]].

\bibitem{Denner:2019vbn}
A.~Denner and S.~Dittmaier,
Phys. Rept. \textbf{864}, 1-163 (2020)
[arXiv:1912.06823 [hep-ph]].

\bibitem{Jegerlehner:2001ca}
F.~Jegerlehner,
[arXiv:hep-ph/0105283 [hep-ph]].

\bibitem{Dittmaier:2009cr}
S.~Dittmaier and M.~Huber,
JHEP \textbf{01}, 060 (2010)
[arXiv:0911.2329 [hep-ph]].

\bibitem{Denner:2006ic}
A.~Denner and S.~Dittmaier,
Nucl. Phys. B Proc. Suppl. \textbf{160}, 22-26 (2006)
[arXiv:hep-ph/0605312 [hep-ph]].

\bibitem{Actis:2006rc}
S.~Actis and G.~Passarino,
Nucl. Phys. B \textbf{777}, 100-156 (2007)
[arXiv:hep-ph/0612124 [hep-ph]].

\bibitem{Frederix:2018nkq}
R.~Frederix, S.~Frixione, V.~Hirschi, D.~Pagani, H.~S.~Shao and M.~Zaro,
JHEP \textbf{07}, 185 (2018)
[erratum: JHEP \textbf{11}, 085 (2021)]
[arXiv:1804.10017 [hep-ph]].

\bibitem{Sirlin:1991fd}
A.~Sirlin,
Phys. Rev. Lett. \textbf{67}, 2127-2130 (1991)

\bibitem{Passarino:2010qk}
G.~Passarino, C.~Sturm and S.~Uccirati,
Nucl. Phys. B \textbf{834}, 77-115 (2010)
[arXiv:1001.3360 [hep-ph]].
\end{thebibliography}
\end{document}